\documentclass[twoside,twocolumn]{article}
\usepackage[super,sort&compress,comma]{natbib} 
\usepackage{mhchem}
\usepackage{times,mathptmx}
\usepackage{sectsty}
\usepackage{balance} 
\usepackage{graphicx}
\usepackage{lastpage}
\usepackage[format=plain,justification=raggedright,singlelinecheck=false,font=small,labelfont=bf,labelsep=space]{caption} 
\usepackage[colorlinks=true,linkcolor=black,citecolor=black,urlcolor=black]{hyperref}
\usepackage{fancyhdr}
\usepackage{rotating}
\usepackage{pdflscape}
\usepackage{multirow}
\usepackage{booktabs}
\usepackage{amssymb}
\usepackage{url}
\usepackage{wrapfig}

\begin{document}

\thispagestyle{plain}
\fancypagestyle{plain}{
\renewcommand{\headrulewidth}{0pt}}
\renewcommand{\thefootnote}{\fnsymbol{footnote}}
\renewcommand\footnoterule{\vspace*{1pt}
\hrule width 3.4in height 0.4pt \vspace*{5pt}} 
\setcounter{secnumdepth}{0}

\renewcommand\floatpagefraction{.99}
\renewcommand\topfraction{.99}
\renewcommand\bottomfraction{.99}
\renewcommand\textfraction{.01}
\renewcommand\dbltopfraction{0.99}
\renewcommand\dblfloatpagefraction{0.99}

\makeatletter 
\def\subsubsection{\@startsection{subsubsection}{3}{10pt}{-1.25ex plus -1ex minus -.1ex}{0ex plus 0ex}{\normalsize\bf}} 
\def\paragraph{\@startsection{paragraph}{4}{10pt}{-1.25ex plus -1ex minus -.1ex}{0ex plus 0ex}{\normalsize\textit}} 
\renewcommand\@biblabel[1]{(#1)}            
\renewcommand\@makefntext[1]
{\noindent\makebox[0pt][r]{\@thefnmark\,}#1}
\makeatother 
\renewcommand{\figurename}{\footnotesize{Figure}~}
\sectionfont{\large}
\subsectionfont{\normalsize} 

\fancyhead{}
\renewcommand{\headrulewidth}{0pt} 
\renewcommand{\footrulewidth}{0pt}
\setlength{\arrayrulewidth}{1pt}
\setlength{\columnsep}{6.5mm}
\setlength\bibsep{1pt}

\twocolumn[
\begin{@twocolumnfalse}
\noindent\LARGE{\bfseries\sffamily Hybrid Perovskites, Metal--Organic Frameworks, and Beyond: Unconventional Degrees of Freedom in Molecular Frameworks}
\vspace{0.3cm}

\noindent\large{{Hanna L. B. Bostr{\"o}m$^{1,2}$ and  Andrew L. Goodwin$^{1,\ast}$}}\vspace{0.3cm}

\noindent\footnotesize{$^1$Department of Chemistry, University of Oxford, Inorganic Chemistry Laboratory, South Parks Road, Oxford OX1 3QR, UK\\
$^2$Max Planck Institute for Solid State Research, Stuttgart, 70569 Germany}

\noindent\rule{\textwidth}{0.4pt}
\noindent \normalsize{{\bf CONSPECTUS:} The structural degrees of freedom of a solid material are the various distortions most straightforwardly activated by external stimuli such as temperature, or pressure, or adsorption. One of the most successful design strategies in materials chemistry involves controlling these individual distortions to produce useful collective functional responses. In a ferroelectric such as lead titanate, for example, the key degree of freedom involves asymmetric displacements of Pb$^{2+}$ cations; it is by coupling these together that the system as a whole interacts with external electric fields. Collective rotations of the polyhedral units in oxide ceramics are another commonly exploited distortion, driving anomalous behaviour such as negative thermal expansion---the counterintuitive phenomenon of volume contraction on heating. An exciting development in the field has been to exploit the interplay between different distortion types: generating polarisation by combining two different polyhedral rotations, for example. In this way, degrees of freedom act as geometric `elements' that can themselves be combined to engineer materials with new and interesting properties. Just as the discovery of new chemical elements quite obviously diversified chemical space, so we might expect that identifying new and different types of structural degrees of freedom to be an important strategy for developing new kinds of functional materials.

In this context, the broad family of molecular frameworks is emerging as an extraordinarily fertile source of new and unanticipated distortion types, the vast majority of which have no parallel in the established families of conventional solid-state chemistry.

Framework materials are solids whose structures are assembled from two fundamental components: nodes and linkers. Quite simply, linkers join the nodes together to form scaffolding-like networks that extend from the atomic to the macroscopic scale. These structures usually contain cavities, which can also accommodate additional ions for charge balance. In the well-established systems---such as lead titanate---node, linker, and extra-framework ions are all individual atoms (Ti, O, and Pb, respectively). But in \emph{molecular} frameworks, at least one of these components is a molecule.

In this Account, we survey the unconventional degrees of freedom introduced through the simple act of replacing atoms by molecules. Our motivation is to understand the role these new distortions play (or might be expected to play) in different materials properties. The various degrees of freedom themselves---unconventional rotational, translational, orientational, and conformational states---are summarised and described in the context of relevant experimental examples. The much-improved prospect for generating emergent functionalities by combining these new distortion types is then discussed. We highlight a number of directions for future research---including the design and application of hierarchically structured phases of matter intermediate to solids and liquid-crystals---which serve to highlight the extraordinary possibilities for this nascent field. \\
}\vspace{-5pt}
\noindent\rule{\textwidth}{0.4pt}

\vspace{0.5cm}
\end{@twocolumnfalse}
]

\section{Introduction}

\begin{figure*} 
\centering
\includegraphics{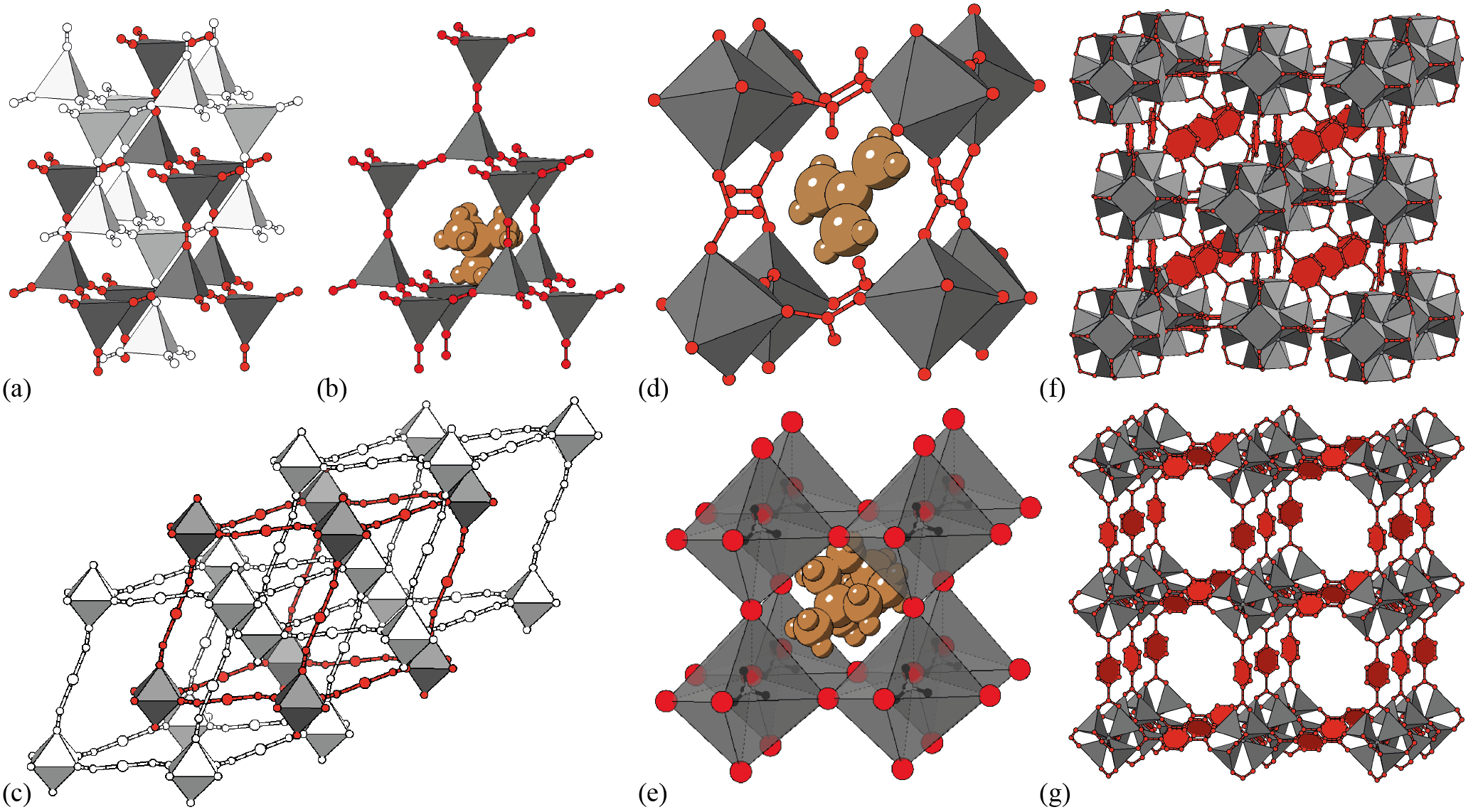}
\caption{Some {\sf A}$_m${\sf BX}$_n$ molecular framework structures, with {\sf A}-site extra-framework ions shown in gold, {\sf B}-site nodes in grey, and {\sf X}-site linkers in red. Interpenetrated framework copies are shown in white. (a) The negative thermal expansion behaviour (NTE) material Zn(CN)$_2$, with its two interpenetrating diamondoid nets.\cite{Zhdanov_1941,Hoskins_1990,Williams_1997,Goodwin2005} (b) In [NMe$_4$]CuZn(CN)$_4$ only one net remains, with [NMe$_4$]$^+$ ions occupying the network cavities.\cite{Hoskins_1990} (c) The three interpenetrating nets of Ag$_3$[Co(CN)$_6$] (= Co[Ag(NC)$_2$]$_3$), a material known for its colossal NTE and strong negative linear compressibility (NLC) effect.\cite{Goodwin2008,Goodwin2008a,Ludi1967a} The hybrid perovskites (d) [Gua][Mn(HCOO)$_3$] (Gua = guanidinium) and (e) [MDABCO][NH$_4$]I$_3$ (MDABCO = methyl-dabconium).\cite{Hu2009,Ye2018} The latter is a metal--free ferroelectric. (f) UiO-66, with chemical formula  [Zr$_6$O$_4$(OH)$_4$][bdc]$_6$ (bdc = 1,4-benzene-dicarboxylate), adopts the {\bf FCU} topology;\cite{Cavka2008} (g) MOF-5, also known as IRMOF-1, has chemical formula [Zn$_4$O][bdc]$_3$ and {\bf PCU} topology.\cite{Li1999} Neither MOF has any cation on the corresponding {\sf A}-sites.}
\label{fig1}
\end{figure*}

Framework materials are a broad and important family of solids that include zeolites, perovskites, metal--organic frameworks (MOFs), and coordination polymers (CPs).\cite{Wells_1984,Hoskins_1990,Yaghi_1998,Cheetham_2007,Ferey_2008,Kitagawa_2004,Horike2009,Li2017} Their network structures are composed of nodes and linkers, and may incorporate counterions and/or neutral guest molecules within their cavities. A very general formula for the family is {\sf A}$_m${\sf BX}$_n\cdot$\{guest\}, where {\sf A} represents the extra-framework counterion(s), {\sf B} the framework node, and {\sf X} the corresponding linker. The famous {\sf ABX}$_3$ stoichiometry of perovskites such as BaTiO$_3$ and MAPbI$_3$ (MA = CH$_3$NH$_3$) is an obvious example;\cite{Megaw1946,Weber1978a,Li2017} the ReO$_3$ networks are related {\sf A}-site-deficient systems ($m=0$) that include \emph{e.g.}\ the archetypal MOF-5, for which {\sf B} = OZn$_4^{6+}$ and {\sf X} is the terephthalate dianion.\cite{Li1999,Evans2020} Some canonical examples are shown in Fig.~\ref{fig1}.


Historically, the task of developing functional framework materials has focussed heavily on purely inorganic systems, such as oxide ceramics.\cite{Goodenough1998,Cheetham_2007} And for good reason: the charge, spin, spin-state, and orbital degrees of freedom of (\emph{e.g.}) transition metals are key for many important physical properties, and the interactions amongst these various components are strongest when frameworks are dense. The incorporation of molecular components necessarily opens up a framework structure---molecules occupy more space than atoms, after all---such that collective properties dependent on the interaction of electronic degrees of freedom generally suffer as a consequence.

The flip-side of this coin is that open framework structures tend to be more flexible.\cite{Tan2011} This flexibility is often cast in terms of \emph{structural} degrees of freedom, a famous example of which is the family of octahedral tilt distortions found in perovskites.\cite{Glazer1972,Woodward1997} From the perspective of exploiting flexibility in terms of functional response, it seems obvious to focus on mechanical properties---and, indeed, many molecular frameworks exhibit unusual mechanical phenomena such as negative thermal expansion (NTE, contraction on heating) and/or negative linear compression (NLC, expansion on compression).\cite{Coates2019,Cairns2015} But a remarkable development has been the realisation that structural degrees of freedom, chosen carefully, can combine to generate electronic degrees of freedom.\cite{Pitcher2015,Oh2015} This is the basis of so-called hybrid improper ferroelectricity, whereby bulk polarisation develops as a consequence of coupling amongst various non-polar structural distortions.\cite{Benedek2011,Benedek2015} Molecular frameworks have more structural degrees of freedom than their conventional counterparts, and so offer many more possibilities for generating emergent physical properties of this type.\cite{Bostrom2018}

In this Account, we survey the key structural degrees of freedom unique to molecular frameworks---`unconventional' in the sense that they are not observed in dense inorganic frameworks. These include `forbidden' tilts, shifts, and orientational and conformational degrees of freedom. We cover each of these aspects in turn, emphasising where possible the scope for chemical control and some of the various functional implications. Before doing so we first summarise briefly the compositional space accessible to these molecular systems. Our review concludes with a forward-looking discussion of opportunities and developments in the field.

\section{Compositions of molecular frameworks}

Molecules can be incorporated within an {\sf A}$_m${\sf BX}$_n$ framework on any one or combination of the {\sf A}-, {\sf B}-, and {\sf X}-sites.

A topical example of molecular substitution on the {\sf A}-site is that of the organic halide perovskites such as MAPbI$_3$, famous for their photovoltaic performance.\cite{Kojima2009,Snaith2018,Correa-Baena2017} We will come to discuss the importance of molecular shape on the structural behaviour and associated functionality of such systems. Whereas single-atom {\sf A}-site cations tend to interact with the surrounding {\sf BX}$_n$ framework predominantly in terms of space-filling (\emph{e.g.}\ the structure-directing tolerance factors) and electrostatic stabilisation, molecular {\sf A}-site cations allow for more chemically complex interactions. A good example is that of hydrogen-bonding between {\sf A}-site guanidinium (Gua$^+$) cations and {\sf X}-site formate anions in guanidinium transition-metal formates: the interactions are sufficiently strong and directional as to hold the [Gua]Fe$_{2/3}\Box_{1/3}$(HCOO)$_3$ ($\Box$ = vacancy) framework together even when one-third of the B sites is absent.\cite{Bostrom2019b} Another important distinction between monoatomic and molecular cations is that of accessible charge states. Molecular cations are predominantly univalent---dabconium, [H--N(C$_2$H$_4$)$_3$N--H]$^{2+}$, being a notable exception\cite{Chen2018b,Wu2017}---whereas inorganic ions can often access higher charges.\cite{Vasala2015} Thus, molecular {\sf A}-site species offer a larger diversity in terms of interactions and shapes, but a smaller range of available charge states, relative to monoatomic {\sf A}-site species. 
 
The scope for compositional variation of the {\sf B}-site is normally system dependent. By way of example, molecular perovskites generally comprise univalent {\sf A} and {\sf X} species, which thus requires {\sf B} to be a divalent octahedral species. As a result, molecular perovskites are frequently based on first-row transition metals---rather than molecular ions---as these satisfy both the charge and geometry requirements.\cite{Li2017} This is in contrast to oxide perovskites, which may feature cations with oxidation states as high as 6+.\cite{Vasala2015}  However, molecular {\sf B}-site cations have been incorporated in hybrid perovskite frameworks: a high-profile example is the metal--free ferroelectric perovskite [MDABCO]NH$_4$I$_3$ (MDABCO = [H--N(C$_2$H$_4$)$_3$N--CH$_3$]$^{2+}$), assembled by H--I interactions.\cite{Ye2018} The concept of molecular nodes is well established in MOF chemistry, with many canonical systems based on such architectures: the [OZn$_4$]$^{6+}$ and [Zr$_6$O$_4$(OH)$_4$]$^{12+}$ {\sf B}-site clusters in MOF-5 and UiO-66, respectively, are obvious examples.\cite{Li1999,Cavka2008} One advantage of incorporating oxometallate clusters is that it allows highly charged metal cations (up to 4+) to be incorporated, which often correlates with good thermal and chemical stability.\cite{Yuan2018,Cavka2008} 

Molecular substitution on the {\sf X}-site leads to the rich families of MOFs and CPs.\cite{Horike2009,Morris2017,Furukawa2013} Organic anions are the predominant type of molecular {\sf X}-site linkers, but metal-containing species may also be used, as in the dicyanometallate\cite{Alexandrov2015,Hill2016} and metalloporphyrin\cite{Gao2014b} frameworks. Even simple inorganic molecules and molecular ions can occupy the {\sf X}-site: examples include BH$_4^-$, SiF$_6^{2-}$, I$_2$, and CN$^-$.\cite{Paskevicius2017,Nugent_2013,Blake1995,Buser1977} Quite obviously the scope for variation on this site is essentially without limit, and includes linker functionalisation---an important tool for property optimisation.\cite{Schneemann2014,Lu2014} For example, the thermal stability,\cite{Marx2010} band gap,\cite{Flage-Larsen2013} and absorption behaviour\cite{Devic2010a} of molecular frameworks can be optimised through judicious choice of {\sf X}-site chemistry. The linker shape may also guide the topology of the system and the propensity for distortions.\cite{Robson2000,Bostrom2020} Hence, of the three site types in molecular frameworks, it is the {\sf X}-site that offers the richest playground for tuning structure and function. 

\section{Conventional and forbidden tilt distortions}

Arguably the best studied type of structural degree of freedom in conventional framework materials is that of tilt distortions. In perovskites, these are the `octahedral tilts' famously categorised by Glazer,\cite{Glazer1972,Howard1998} and more generally they are the rigid-unit modes (RUMs) relevant to \emph{e.g.}\ the silicate minerals and zeolites.\cite{Hammonds1996,Dove1995,Dove1997} Their ubiquity arises from the contrast in energy scales between the cost of deforming the tightly bound {\sf BX}$_{2n}$ coordination polyhedra and flexing the underconstrained {\sf B}--{\sf X}--{\sf B} linkages. As a consequence, a common structural distortion involves correlated rotations of corner-sharing coordination polyhedra, themselves behaving effectively as rigid bodies. By way of example, the famous tilt instability in SrTiO$_3$ involves counter-rotation of corner-sharing TiO$_6$ octahedra around a common axis---the Ti coordination geometry is preserved in the process, but the Ti--O--Ti angle flexes to reduce the system volume at low temperatures.\cite{Cowley_1964,Yamada1969,Lytle1964} In general, many different tilt distortions are possible for a given framework, with each one breaking crystal symmetry in its own particular way. This symmetry breaking can be exploited in the design of hybrid improper ferroelectrics, which has led to the development of so-called `tilt engineering' approaches,\cite{Pitcher2015,Li2015} whereby framework composition is cleverly manipulated to introduce tilt distortions of a specific useful symmetry.

\begin{figure} 
\centering
\includegraphics{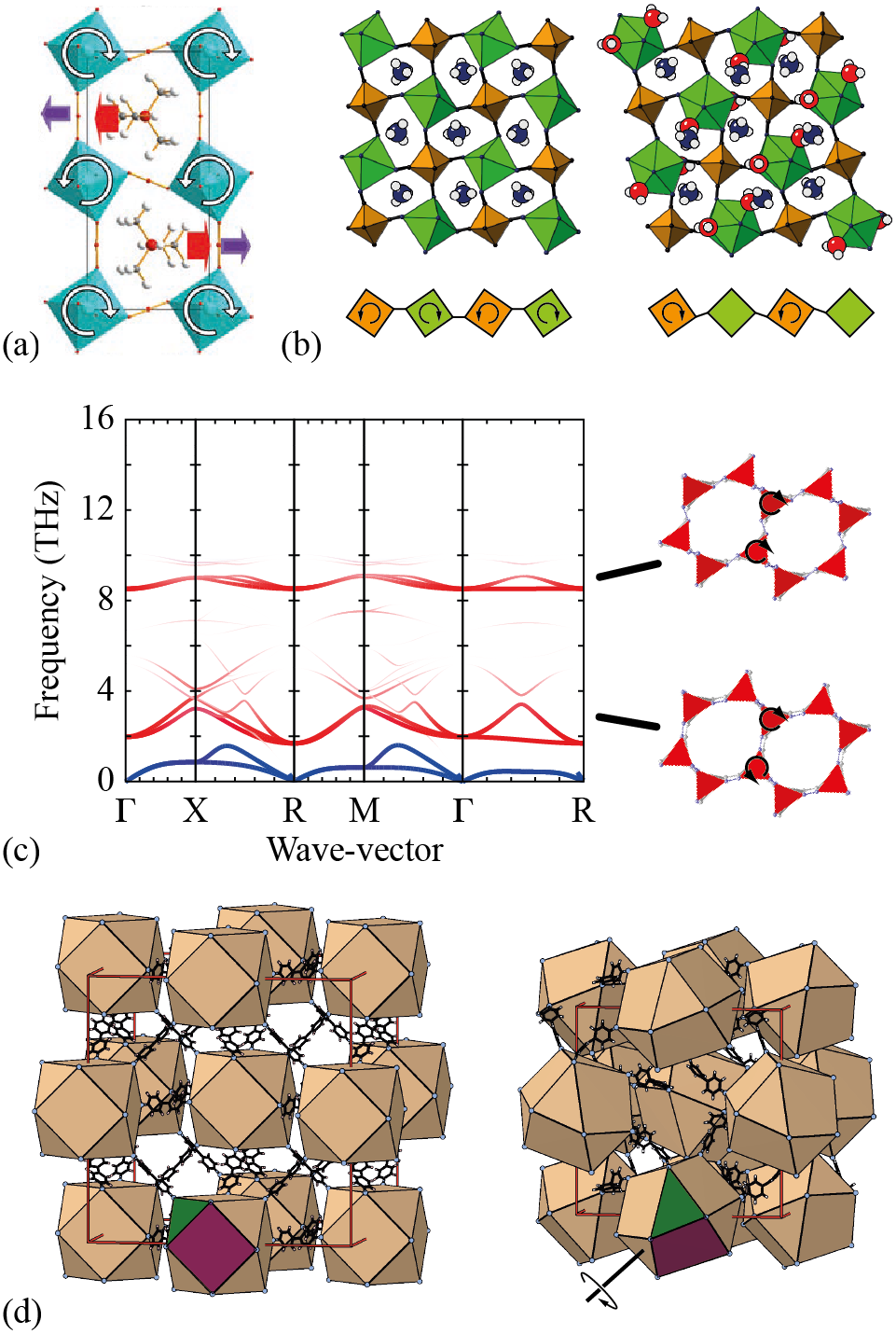}
\caption{Tilt degrees of freedom in molecular frameworks. (a) Octahedral tilts in [NMe$_4$]Mn(N$_3$)$_3$ involve a combination of conventional (out-of-phase) and forbidden (in-phase) rotations. Figure reproduced from Ref.~\citenum{Gomez-Aguirre2016}. (b) The octahedral tilt system in [NH$_4$]SrFe(CN)$_6$ can be switched between conventional (left) and forbidden (right) configurations through reversible (de)hydration at the Sr centres. Figure adapted from Ref.~\citenum{Duyker2016}. (c) The low-energy phonon dispersion curves of Zn(CN)$_2$ include essentially dispersionless branches associated with Zn-tetrahedral tilts (shaded in red); these partition into modes involving translations (2--4\,THz) and rotations (8--10\,THz) of the {\sf X}-site cyanide ions. Figure adapted from Ref.~\citenum{Fang2013a}. (d) The phenomenon of NGA in DUT-49 involves a reversible transition between open (left) and dense (right) states. These two states, which differ in molar volume by a factor of two, are related by a large-magnitude collective rotation of the {\sf B}-site polyhedra around $\langle111\rangle$ axes of the cubic unit cell. The flexibility of the polyphenyl {\sf X}-site linker is key to allowing this transformation.\cite{Krause2016}}
\label{fig2}
\end{figure}

An important distinction between atomic and molecular {\sf X}-site linkers is that, loosely speaking, the former requires neighbouring [{\sf BX}$_{2n}$] polyhedra to rotate in opposite directions but the latter lifts any such constraint.\cite{Goodwin2006} In particular, it is possible in molecular frameworks for neighbouring polyhedra to rotate in the same sense as one another---a so-called `forbidden' tilt distortion that has no analogue in conventional frameworks.\cite{Duyker2016} It turns out that forbidden tilts are a relatively common phenomenon in many molecular perovskites, with a particular predominance in azide-bridged systems [Fig.~\ref{fig2}(a)].\cite{Bostrom2020,Zhao2013,Du2015a}  In the Prussian blue-like framework [NH$_4$]$_2$SrFe(CN)$_6$, it is even possible to switch between conventional and forbidden tilt distortions through reversible guest binding at the {\sf B}-site [Fig.~\ref{fig2}(b)].\cite{Duyker2016}

The accessibility of forbidden tilts to frameworks with molecular {\sf X}-sites has two clear functional implications. The first is the profound increase in diversity of symmetry-breaking distortions one might introduce, which in turn expands the possibility for tilt engineering.\cite{Bostrom2018} The second is an increased density of low-energy volume-reducing tilt modes in the vibrational spectrum of these systems. Whereas conventional tilt modes are localised at the Brillouin zone boundary---reflecting the alternation in rotation sense from polyhedron to polyhedron---forbidden tilts can usually be associated with all possible wave-vectors (\emph{i.e.}\ periodicities), and so their contribution to macroscopic thermodynamic properties is more substantial.\cite{Goodwin2006} This point is thought to help explain why many molecular frameworks exhibit strong NTE effects: on heating, one populates the whole family of volume-reducing tilt modes, which is sufficient to overcome the usual positive thermal expansion contribution from other vibrations.\cite{Goodwin2006,Coates2019} This point is nicely illustrated in the case of the isotropic NTE material Zn(CN)$_2$, for which the phonon spectrum is well understood [Fig.~\ref{fig2}(c)].\cite{Fang2013a,Goodwin2005}

As the length of an {\sf X}-site anion increases, so too does its capacity to allow for extreme flexing in response to external stimuli. A topical example is the phenomenology of negative gas adsorption (NGA), as observed in the Cu-based MOF DUT-49.\cite{Krause2016} The effect itself is rather bizarre: on exposure to increasing gas pressure, crystals of DUT-49 first expand as the gas fills their pores, before collapsing at a critical pressure and releasing the included gas molecules (under pressure!). The transition between open and dense states is driven by a collective tilt mode of large multi-centre nodes, facilitated by extreme flexing of polyphenyl {\sf X}-site linkers [Fig.~\ref{fig2}(d)]. Although the tilt itself turns out to be conventional, rather than forbidden, its magnitude is entirely unconventional---indeed sufficient to drive a volume collapse greater than 50\%.

\section{Columnar shifts, breathing, and loop moves}

\begin{figure} 
\centering
\includegraphics{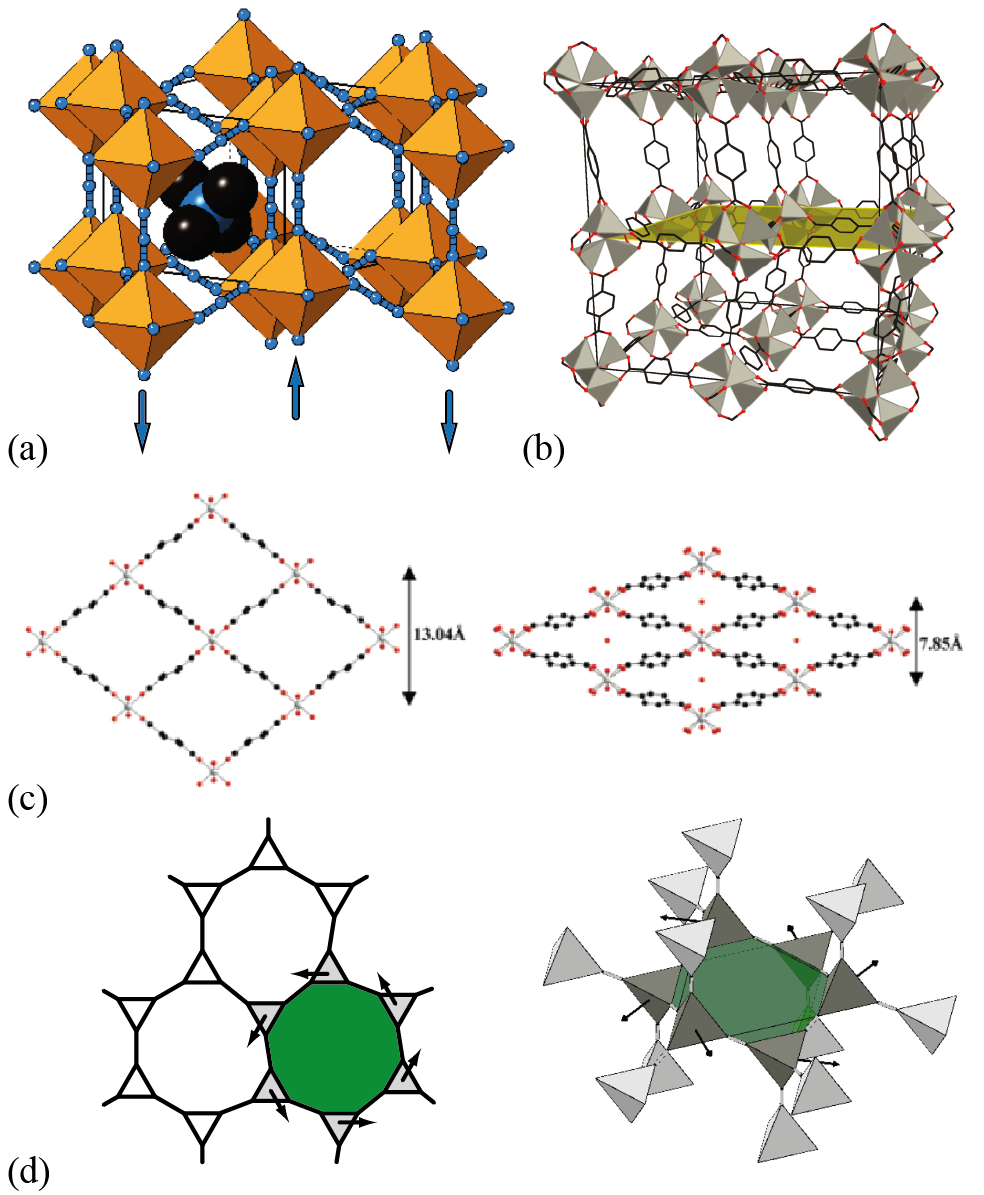}
\caption{Translational degrees of freedom in molecular frameworks. (a) The columnar shifts of hybrid perovskites involve translations of columns or planes of connected polyhedra, as shown here for the case of [(CH$_3$)$_4$N]Ca(N$_3$)$_3$.\cite{Mautner1988,Bostrom2016} Note the coupling between shift displacement pattern and {\sf A}-site molecular shape. Figure adapted from Ref.~\citenum{Bostrom2016}. (b) The low shear modulus of MOF-5 is understandable in terms of low-energy planar shift degrees of freedom. Figure reproduced from Ref.~\citenum{Rimmer2014}. (c) The large-pore/narrow-pore transition of MIL-53, involves adsorption-driven framework shear. Figure adapted from Ref.~\citenum{Serre2002}. (d) In some molecular frameworks, such as those with the augmented kagome (left) or augmented pyrochlore (right) nets, localised multi-node translational degrees of freedom exist that correspond to collective rotations of large structural units (green regions). These emergent degrees of freedom are termed `loop moves'. Figure adapted from Ref.~\citenum{Goodwin2006}.}
\label{fig3}
\end{figure}

Molecular linkers also enable new low-energy deformations involving collective \emph{translations} that are impossible in inorganic frameworks.\cite{Goodwin2006} These modes are most straightforwardly understood in molecular perovskites, where they are referred to as `columnar shifts'.\cite{Bostrom2016} They often play an important symmetry-breaking role in these systems and are well placed to drive hybrid improper ferroelectricity.\cite{Bostrom2016,Bostrom2018,Chen2014a} As for unconventional tilts, columnar shifts are frequently observed in frameworks with azide or dicyanamide ligands [Fig.~\ref{fig3}(a)].\cite{Zhao2013,Bostrom2020}

The asymmetric naphthalenedicarboxylate (ndc) linker induces a similar effect in the MOF known as DUT-8(Ni). Here, neighbouring columns of dabco-bridged nickel--carboxylate paddlewheels are shifted relative to one another as a consequence of the step-like ndc geometry.\cite{Petkov_2019} Whereas in hybrid perovskites the shifts are usually well ordered, in DUT-8(Ni) they exhibit a strongly-correlated disorder governed by strict local rules: each square pore is bounded by two ndc step-ups and two step-downs.\cite{Reynolds_2020} The particular type of disordered arrangement can even be controlled reversibly by guest (de)sorption, which imparts the system with an unusual type of structural flexibility.\cite{Ehrling_2020}

In the long-wavelength limit, shift modes correspond to a shear distortion of the framework structure. Hence the propensity for molecular frameworks with cubic or square topologies to exhibit shift-type instabilities is reflected also in their low shear moduli---as in MOF-5 (Refs.~\citenum{Mattesini_2006,Rimmer2014})---and \emph{e.g.}\ the existence of ferroelastic open-pore/narrow-pore transitions [Fig.~\ref{fig3}(b,c)]. Such transitions are known as `breathing modes' in flexible MOFs.\cite{Serre2002,Serre_2007b,Schneemann2014,Morris2017,Henke2012} Perhaps the best known example is MIL-53, where hydration leads to a winerack-type contraction of the framework due to the hydrogen bonding interactions between guest water and the terephthalate linkers.\cite{Serre2002} Conversely, the isoreticular series of MIL-88 exhibits a remarkable \textit{expansion} upon guest absorption, with a volume change of up to 300\%.\cite{Serre2007} The breathing ability renders these MOFs suitable candidates for diverse applications such as drug delivery,\cite{Horcajada2008} removal of hazardous materials,\cite{Khan2013} and gas separation.\cite{Carrington2017} In addition to sorption-induced strain, winerack-type hingeing can sometimes be triggered by pressure, leading to the rare and counterintuitive phenomenon of negative linear compressibility---expansion in one or two directions upon the application of hydrostatic pressure.\cite{Cairns2012,Li_2012,Cairns2015,Ortiz2012} This has been observed in several MOFs and CPs, including MIL-53 and the triply interpenetrated Prussian blue-like Ag$_3$[Co(CN)$_6$].\cite{Serra-Crespo2015,Goodwin2008a}

Whereas translational degrees of freedom in cubic and square network topologies are well understood, the case is much less clear for other systems. For molecular frameworks with diamondoid topologies, a very different type of collective translation occurs: hexagonal loops of connected tetrahedral [{\sf BX}$_4$] units rotate such that each tetrahedron translates along a loop tangent [Fig.~\ref{fig3}(d)].\cite{Goodwin2006} We term these collective displacements `loop moves', by analogy to the collective loop spin degrees of freedom in Ising pyrochlore magnets.\cite{Jaubert_2011,Melko_2001,Melko_2004}  A peculiarity of the diamond network is that there are exactly as many loops as tetrahedral nodes, which has the unexpected consequence that each {\sf BX}$_4$ unit retains a full translational degree of freedom.\cite{Goodwin2006} Polyhedral displacements are also observed in the quartz-structured molecular framework $\alpha$-Zn[Au(CN)$_2$]$_2$, where a pressure-induced phase transition is accompanied by a coupled translation of the [ZnN$_4$] tetrahedra.\cite{Cairns2013a} The extension from these select few examples to a general understanding of translational degrees of freedom in molecular frameworks of arbitrary topology, however, remains very much an open question and work-in-progress.

\section{Orientational and multipolar order}

An obvious difference between inorganic and molecular species is the aspherical symmetry of the latter. This distinction can be exploited in the search for non-centrosymmetric structures. Curie's principle states that a crystal will adopt the common symmetry subgroup of the point symmetries of its components. Hence, non-centrosymmetric structures can be designed by judicious choice of components with specific symmetries.\cite{Ye2018} Moreover, the anisotropy inherent to molecular species allows for orientational degrees of freedom that simply do not exist in conventional frameworks.

\begin{figure} 
\centering
\includegraphics{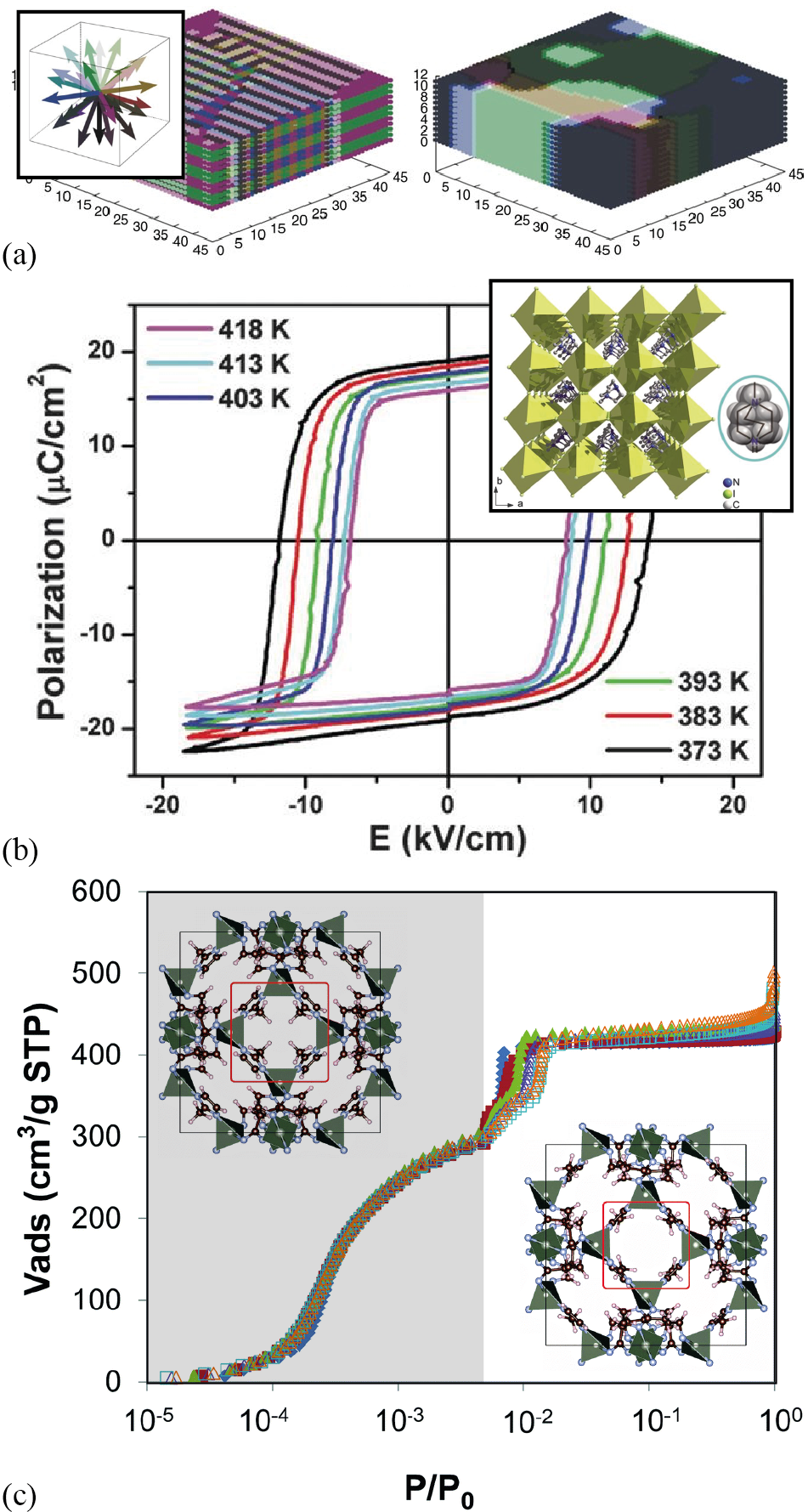}
\caption{Nature and consequences of orientational degrees of freedom in some molecular frameworks. (a) In MAPbI$_3$, the {\sf A}-site methylammonium cation is free in principle to orient along a number of directions within the perovskite cage (coloured arrows). The combination of dipolar interactions with either antiferroelectric (left) or ferroelectric (right) couplings leads to polarisation textures that influence exciton dynamics. Figure adapted from Ref.~\citenum{Leguy2015}. (b) In the metal-free ferroelectric [MDABCO][NH$_4$]I$_3$, bulk polarisation emerges from collective orientational order of the polar [MDABCO]$^{2+}$ cations. The polarisation can be flipped in an applied electric field. Figure adapted from Ref.~\citenum{Ye2018}. (c) Under application of a critical gas pressure, the imidazolate linkers in ZIF-8 switch orientations to increase the pore window size of its sodalite cage structure (the `swing effect'). The sorption profiles within closed-pore (grey) and open-pore (white) configurations differ; the latter is particularly dependent on particle size (varying coloured data). Figure adapted from Refs.~\citenum{Tian2016,Casco_2016}.}
\label{fig4}
\end{figure}

Several properties of molecular perovskites depend on orientational order and disorder of the {\sf A}-site molecule. For example, the dynamics of the methylammonium cations in the photovoltaic material MAPbI$_3$ affects exciton lifetimes, which is crucial for its performance in solar cells [Fig.~\ref{fig4}(a)].\cite{Leguy2015,Zhu2016} Furthermore, cyanoelpasolites---{\sf A}$_2${\sf B}[{\sf B}$^\prime$(CN)$_6$], where {\sf A} is an organic cation---typically display phase transitions upon cooling, driven by progressive freezing of the motion of the {\sf A}-site cation. This can be exploited for the development of materials with switchable dielectric constants.\cite{Zhang2010,Xu2016a,Zhang2013} The metal-free ferroelectric [MDABCO][NH$_4$]I$_3$ develops a polarisation competitive with that in BaTiO$_3$ through orientational order of the polar [MDABCO]$^{2+}$ cations [Fig.~\ref{fig4}(b)].\cite{Ye2018}

Phase transitions relating to orientational order can often be described in terms of multipole moments and map onto relatively simple physical models.\cite{Parsonage_1978,Ziman_1979,Simenas2016,Simenas2019a,Coates2019a,Simonov_2020} For example, the onset of rotation of the imidazolium cation in (H$_2$Im)$_2$K[Fe(CN)$_6$] corresponds to loss of dipolar order, whereas---as the normal of the imidazolium plane is unchanged---the quadrupolar order is retained.\cite{Zhang2010} The type of multipolar order is dictated by the symmetry relationship between the point symmetry of the {\sf A}-site cation and its site in the ideal undistorted parent.\cite{Coates2019a} Some control over the order of the multipole moments may be achieved by considering the packing efficiency and size of the cavity.\cite{Evans2016} As a result, design rules based on multipole moments may be within reach. An interesting parallel to these examples is the phenomenology of so-called `hidden-order transitions' in strongly-correlated electronic materials, \textit{e.g.} the heavy-fermion system URu$_2$Si$_2$, which also involve the emergence of multipolar order on lattices of various topologies.\cite{Paddison2015,Ikeda2012} 

If the symmetry of the {\sf B}-site species is lower than the point symmetry of its crystallographic site, then orientational {\sf B}-site order can emerge. A well-known example in conventional frameworks is that of collective Jahn-Teller order---crucial to the physics of many strongly-correlated oxides\cite{Goodenough_1955,Goodenough1998,Tokura2000}. Related degrees of freedom exist in molecular frameworks. For example, the hexagonal Zr-MOF PCN-223 features a Zr oxyhydroxide cluster with three-fold orientational disorder at the nodes of the hexagonal lattice.\cite{Feng2014} Such disorder can complicate the structure determination, as recently noted for PCN-221---a polymorph of PCN-223.\cite{Koschnick2021} The unusual cubic Zr oxyhydroxide cluster initially reported in PCN-221 appears to be the result of a superposition of statically disordered octahedral clusters.\cite{Feng2013,Koschnick2021}

Turning to the {\sf X}-site, the use of cylindrically asymmetric anions leads to potential symmetry breaking by rotation around the {\sf B}--{\sf X}--{\sf B} linkage. This is of particular currency for ZIFs, and has been extensively studied in the specific case of ZIF-8, a porous material with the \textbf{SOD} topology.\cite{Sung2006} It readily absorbs molecules larger than the size of the pore window,\cite{Huang2006} which can be attributed to a rotation of the surrounding imidazolate linker edges (the so-called `swing effect') [Fig.~\ref{fig4}(c)].\cite{Moggach2009a,Fairen-Jimenez2011a} Critically, the degree of rotation can be tuned by chemical functionalisation, which has obvious implications for applications within gas storage and separation.\cite{Hobday2018} Similar gate-opening effects exist for a number of other MOFs.\cite{Gucuyener2010a,Wharmby2015,Hyun2016} Furthermore, a variable-temperature study on cristobalite-like Cd(Im)$_2$ (Im = imidazolate) revealed reorientation of the imidazolate linker, causing an anisotropic and nonlinear thermal response.\cite{Collings2013} Finally, order/disorder processes of the polar dicyanamide ligand, N(CN)$_2^-$, likely contribute to the dielectric anomaly observed on heating in the perovskite [NPr$_4$]Mn[N(CN)$_2$]$_3$ (Pr = C$_3$H$_7$).\cite{Bermudez-Garcia2015} Part of the interest in these systems arise from the possibility of controlling order \textit{via} the application of external electric fields.\cite{Knebel2017}


\section{Molecular conformations}

As the structural complexity of molecular components increases, so too does their capacity for internal degrees of freedom. A simple example is the conformational state (dihedral angle) of the methylammonium cation, which may influence exciton recombination rates in MAPbI$_3$.\cite{Zhu2016} More complex is the case of bis(trisphenylphosphine)iminium (PPN), which acts as the {\sf A}-site cation in dicyanometallate frameworks and which can adopt multiple conformations in the solid state.\cite{Lefebvre2007,Hill2016} When enclosed in the layered [PPN]$_{0.5}$Cd[Ag(CN)$_2$]$_{2.5}$(EtOH), for example, the phenyl groups are oriented as to maximise the $\pi$-$\pi$ interaction between the aromatic rings and the framework. By contrast, in the perovskite [PPN]Cd[Au(CN)$_2$]$_3$, inter- and intramolecular interactions are favoured.\cite{Hill2016} Likewise, internal rotational degrees of freedom distinguish the polymorphic Zr-MOFs NU-1000 and NU-901.\cite{Mondloch2013,Kung2013,Webber2017} These are chemically identical, but topologically inequivalent, and the particular polymorph is dictated by the rotational state of a common pyrene-based ligand. The reaction selectivity can be guided by judicious choice of modulator.\cite{Webber2017,Garibay2018} The polymorphs deviate in their optical properties, as the conformation of the ligand determines the propensity for excimer formation of the chromophoric pyrene linker.\cite{Yu2018}

\begin{figure} 
\centering
\includegraphics{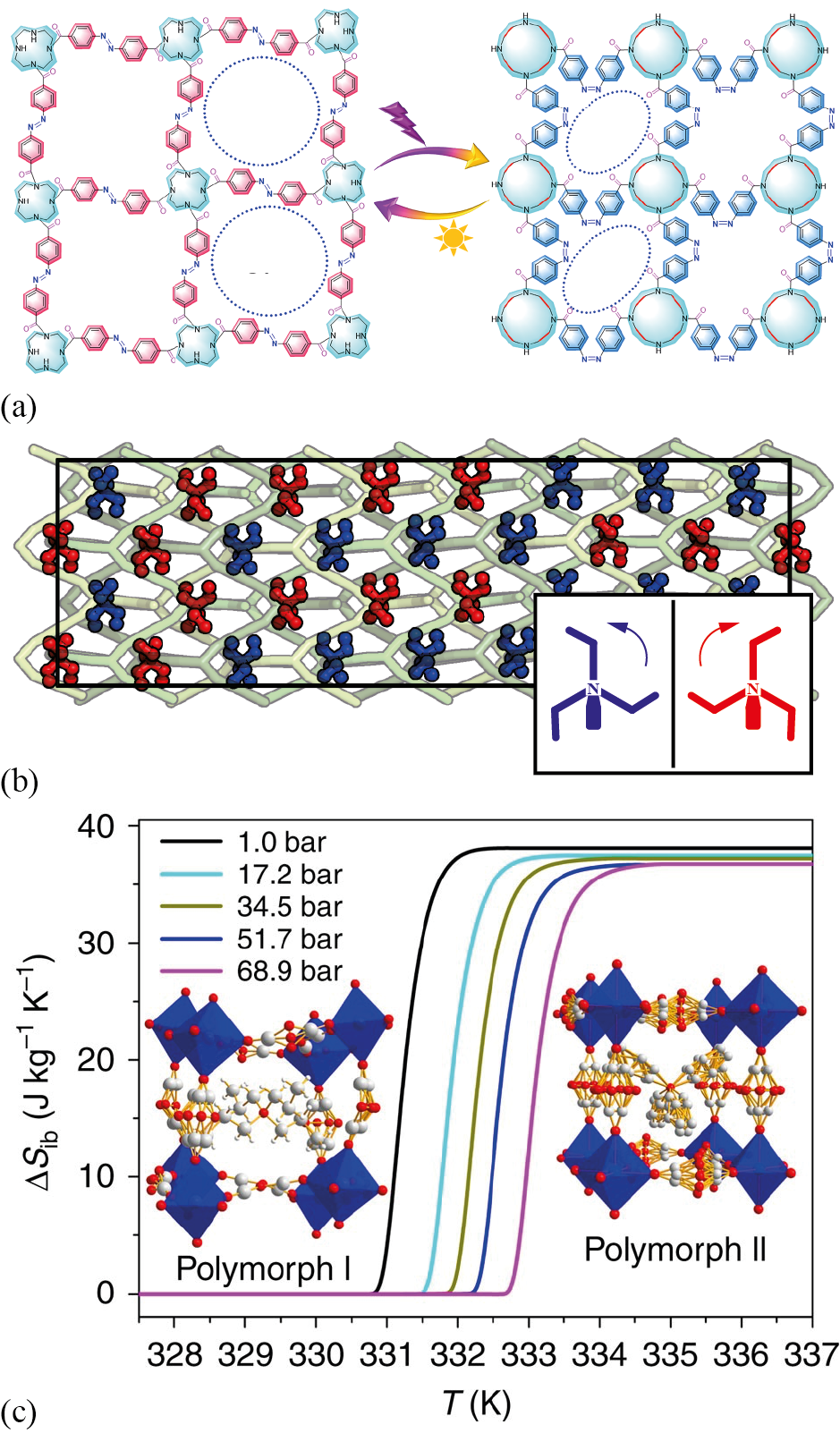}
\caption{Conformational degrees of freedom in molecular frameworks. (a) Reversible light-induced \emph{cis}/\emph{trans} isomerism in an azobenzene-containing covalent organic framework allows control over pore size and shape. Figure adapted from Ref.~\citenum{Liu2020c}. (b) When contained within the cavities of the \{Ag[Ag(CN)$_2$]$_2$\}$^-$ network (transparent green), the otherwise achiral NEt$_4^+$ cations adopt one of two equivalent enantiomeric conformations (shown here in blue and red). Competition between conformational ordering and a shear instability of the lattice leads to an unusual incommensurate chirality density wave (XDW) state, where chirality is modulated over the nanometre lengthscale. Figure adapted from Ref.~\citenum{Hill2017}. (c) Entropy change associated with the conformational order/disorder (I/II) transition in [NPr$_4$]Mn[N(CN)$_2$]$_3$. The transition is suppressed by application of hydrostatic pressure, which provides a strategy for barocaloric cooling. Figure reproduced from Ref.~\citenum{Bermudez-Garcia2017}.}
\label{fig5}
\end{figure}

If molecular conformations can be interchanged by application of external stimuli, then it is possible to exploit such internal, conformational, degrees of freedom in a functional sense. In azobenzene-based porous covalent organic frameworks, for example, irradiation by ultraviolet light leads to reversible \emph{cis}/\emph{trans} isomerism of the linkers.\cite{Liu2020c} This structural change tightens the pore size, which effectively switches on and off the material's ability to transport molecules beyond a certain critical size [Fig.~\ref{fig5}(a)]. A related effect occurs in the cyclohexane-bridged MOF-5 analogue ZrCDC, where desorption-driven chair/boat conformation inversion results in a reversible transition between crystalline and amorphous states.\cite{Bueken2016} In the dicyanometallate-linked diamondoid framework [NEt$_4$]Ag[Ag(CN)$_2$]$_2$ (Et = C$_2$H$_5$), the normally achiral NEt$_4^+$ {\sf A}-site cation adopts a chiral conformation through its interaction with the dicyanometallate network.\cite{Hill2017} At low temperatures this chirality is coupled in a complex fashion throughout the crystal to give an incommensurately modulated `chirality density wave' of potential interest in advanced photonics [Fig.~\ref{fig5}(b)]. On heating, however, the system is eventually able to overcome the interconversion barrier between enantiomorphic conformations and an achiral state emerges. So, again, an external stimulus---in this case temperature---can switch on and off a physical property dependent on molecular degrees of freedom.


An appealing variation on this theme is the exploitation of conformational order/disorder transitions in the design of barocalorics, for which the high-profile example is  [NPr$_4$]Mn[N(CN)$_2$]$_3$.\cite{Bermudez-Garcia2017,Bermudez-Garcia2017b} On heating above a critical temperature of 331\,K, the tetrapropylammonium {\sf A}-site cations switch from being conformationally ordered to a disordered state. A significant entropy change accompanies the transition. Because the disordered state has a larger molar volume, the transition can be suppressed by applying a very modest hydrostatic pressure [Fig.~\ref{fig5}(c)]. A solid-state cooling strategy follows naturally. Starting from the disordered (high temperature) state, pressure is applied until the {\sf A}-site cations order; the entropy loss is expelled as waste heat. As pressure is released, the system disorders, taking in heat (\emph{i.e.} cooling its environment) in order to provide the necessary entropy gain.

\section{Future directions}

For each of the types of degree of freedom and various applications discussed above, there is obvious need and scope for developing clear strategies for chemical control and performance optimisation. This is actually an enormous challenge, given the chemical and structural diversity of the broad family of molecular frameworks. There will also be many other degrees of freedom not covered in this brief Account that are nonetheless worthy of exploration. Examples include the twist-modes of paddlewheel units,\cite{Wu_2008} protonation states of organic linkers,\cite{Umeyama_2012} binding modes,\cite{Bostrom2020} rotational degrees of freedom of motor-like components,\cite{Evans_2021} and the topological degrees of freedom associated with bond rearrangements;\cite{Bennett_2010,Hunt2015} there will be many others.

Just as the manganites have assumed a special role in device physics because they combine many interacting degrees of freedom---charge, spin, lattice, and orbital---so is it the case that we expect the \emph{interplay} of degrees of freedom in molecular frameworks to provide a rich source of future discoveries. Already there are key signs: an obvious example is the complex interplay of framework distortions and molecular reorientations in the exciton physics of MAPbI$_3$.\cite{Frost_2014,Herz_2018} In a similar vein lies the development of clear rules for combining different distortions of molecular perovskites to engineer specific properties, with the case of emergent polarisation in [Gua]Cu(HCOO)$_3$ as a result of combined Jahn-Teller and multipolar order as an excellent example.\cite{Stroppa2011,Evans2016} Solid-solution chemistry---termed `multivariate' synthesis in the MOF field---offers a surprisingly underexplored additional dimension for exploration in molecular framework materials design.\cite{Chen2020a,Yuan2015} Here the scope is especially broad, since substitution of molecular components can involve varying not only size or charge, but also shape, or conformation, or rigidity, or functionality.

A necessary consequence of the larger physical separation between transition-metal centres in magnetic molecular frameworks is a reduced energy scale associated with collective magnetic order. And while examples of strong\cite{Verdaguer_2005} and unconventional\cite{Harcombe_2016} magnetism can be found amongst this broader family of materials, one expects that conventional inorganic or intermetallic systems will always have the upper hand in this regard. Nevertheless the orientational degrees of freedom of \emph{e.g.}\ {\sf A}-site cations can behave as pseudospins that, in favourable cases, interact in a manner analogous to various types of magnetic exchange with strengths $\sim$100\,K.\cite{Coates2019a} Hence there is enormous scope to exploit the way in which molecular frameworks both organise pseudospin degrees of freedom and control their interactions to access structural analogues of exotic magnetic phases.\cite{Simonov_2020}

One target, for example, is the realisation of skyrmionic phases of potential application in information storage and manipulation.\cite{Jonietz_2010,Romming_2013} In such systems the director field associated with {\sf A}-site orientations exhibits the same topological features as in nematic liquid crystals,\cite{Leonov_2014,Wolpert_2020} now driven by the Dzyalonshinskii-Moriya physics of skyrmionic magnets.\cite{Rossler_2006,Yi_2009} Given the extraordinary diversity of structural degrees of freedom found in molecular frameworks, and the varied types of interaction that exist between those degrees of freedom, one anticipates the discovery of all sorts of emergent states (like skyrmions) in molecular frameworks, well beyond those found in the conventional playgrounds of unconventional physics.

\subsection*{Corresponding author}
andrew.goodwin@chem.ox.ac.uk

\section*{Acknowledgements}
The authors gratefully acknowledge financial support from the E.R.C (Grant 788144), the E.P.S.R.C. (Grant EP/G004528/2). Joshua Hill (UCL) is acknowledged for useful discussion.

\renewcommand\refname{References}

\balance

\footnotesize{
\bibliography{refs}
\bibliographystyle{rsc}
}

\end{document}